# Scaling of the specific heat in superfluid films

Norbert Schultka and Efstratios Manousakis
*Department of Physics, Florida State University, Tallahassee, Florida 32306*

We study the specific heat of the $x-y$ model on lattices $L \times L \times H$ with $L \gg H$ (i.e. on lattices representing a film geometry) using the Cluster Monte–Carlo method. In the $H$–direction we apply Dirichlet boundary conditions so that the order parameter in the top and bottom layers is zero. We find that our results for the specific heat of various thickness size $H$ collapse on the same universal scaling function. The extracted scaling function of the specific heat is in good agreement with the experimentally determined universal scaling function using no free parameters.

The theory of second order phase transitions is based on the assumption that at temperatures close to the critical temperature $T_c$ there is only one dominating length scale associated with the critical behavior of the system, the correlation length. Because the correlation length diverges as the critical temperature is approached the microscopic details of the system become irrelevant for the exponents describing the most singular dependence of the thermodynamic functions on the reduced temperature $t = T/T_c - 1$. This intuitive picture has its foundation in the renormalization group treatment of second order phase transitions. Within the renormalization group treatment it becomes evident that the critical behavior can be divided into different universality classes which are characterized by a set of critical exponents.

If the system is confined in a finite geometry (e.g. a cubic or film geometry) the finite–size scaling theory [1] is thought to describe well the behavior of the system at temperatures near $T_c$. The intuitive idea behind the finite–size scaling theory is that finite–size effects are observed when the bulk correlation length $\xi$ becomes of the order of the finite system size, i.e. for our case here the film thickness $H$. For a physical quantity $O$ this statement can be expressed as follows [2]:

$$\frac{O(t,H)}{O(t,H=\infty)} = f\left(\frac{H}{\xi(t,H=\infty)}\right). \qquad (1)$$

The dimensionless ratio on the left-hand-side of the above equation is a universal function $f(x)$ of the dimensionless ratio $x = H/\xi$, i.e. in our present work the film thickness measured with respect to the correlation length. The function $f$ depends on the boundary conditions and the geometry of the system.

Liquid helium $^4He$ has been a very good ground both for testing finite-size scaling theory and measuring the critical exponents that go along with a second order phase transition in the case of a complex order parameter. Measurements of the superfluid density [3] and specific heat [4] on helium films, however, fail to verify the finite-size scaling theory. Field theoretical calculations for the standard Landau–Ginzburg free energy functional in different geometries with Dirichlet boundary conditions have been carried out [5–8]. New specific heat measurements [9] and also a reanalysis [10] of the old specific heat data [4] show good agreement between the results of the calculations reported in Refs. [5–7] and those data. Furthermore, new experiments on liquid $^4He$ under microgravity conditions are planned [11] to examine the finite–size scaling properties of the specific heat. The above mentioned field theoretical calculations are based on methods such as a loop expansion which systematically includes corrections due to fluctuations around a mean-field solution. These methods are approximate in nature and assume that the basic physics of the mean field treatment is qualitatively correct. In addition these methods neglect the effect of vortices. Thus, accurate numerical calculations of the specific heat of helium films and its scaling with film thickness are welcome.

In our previous work, we used the $x-y$ model with periodic boundary conditions [12,13] in the direction of the film thickness $H$ to compute the superfluid density and the specific heat of thin helium films. We demonstrated scaling with the expected values for the critical exponents of the superfluid density and the specific heat with respect to the film thickness, thus confirming the validity of the finite–size scaling theory. However, the obtained universal function for the specific heat does not match the experimentally determined universal function of Ref. [9], indicating that periodic boundary conditions are only a poor approximation of the correct physical boundary conditions.

In this paper we study the effect of Dirichlet boundary conditions on the finite–size scaling behavior of the specific heat of $^4He$ in a film geometry $L \times L \times H$ with $L \gg H$. These boundary conditions are believed to simulate the boundary conditions of the real helium films more accurately [6,7]. With these boundary conditions, the superfluid order parameter vanishes in the top and the bottom of the film. The specific heat of helium films has been measured in films with very large planar dimensions. The large planar dimension ($L$) of the films is taken into account in our calculations by assuming periodic boundary conditions in the planar $L$–directions and $L \gg H$ so that the finite-size effects on our calculated specific heat due to $L$ are not important. We find that our results for the specific heat obey the finite-size scaling theory using the expected values of the critical exponent $\nu$. More importantly, our calculated universal scaling function agrees well with that obtained by the recent measuments of the specific heat of helium films [9] with no adjustable parameters.

We use the $x-y$ model to describe the fluctuations of the order parameter in superfluid $^4He$ near the $\lambda$–critical point (cf. e.g. Ref. [14]). In the pseudospin notation, the $x-y$ model takes the following form:

$$\mathcal{H} = -J \sum_{\langle i,j \rangle} \vec{s}_i \cdot \vec{s}_j, \tag{2}$$

where the summation is over all nearest neighbors, $\vec{s} = (\cos\theta, \sin\theta)$, and $J$ sets the energy scale. The angle $\theta$ corresponds to the phase of the superfluid order parameter $\psi(\vec{r})$. This order parameter is the average value of the helium atom creation operator which is defined in a volume whose linear extensions are much larger than the interparticle spacing and much smaller than the correlation length. This condition can be realized only very near the critical temperature.

In our calculations Dirichlet boundary conditions can be imposed by coupling the top and bottom layers of the film to two *static* staggered pseudospin configurations playing the role of the "substrate" layers so that the pseudomagnetization (the superfluid order parameter) in the top and bottom substrates is exactly zero, i.e.

$$\vec{s}(x,y,z) = (-1)^{x+y}\, \vec{s}(1,1,z), \quad z = 0,\ H+1, \tag{3}$$

where $x$ and $y$ label the integer coordinates of the lattice sites in the two "substrate" planes perpendicular to the $H$–direction one below the film, i.e. $z=0$, and one above the film, i.e. $z = H+1$. In the $L$–directions we applied periodic boudary conditions. The crucial difference between the boundary conditions used in this work and periodic boundary conditions in all directions [12,13] is that the superfluid density develops a profile in the $H$–direction, whereas it is completely homogeneous for periodic boundary conditions.

We computed the specific heat on $L^2 \times H$ size lattices, where $L = 20, 40, 60, 80, 100$ and $H = 12, 16, 20, 24$. The specific heat $c$ is obtained by

$$c = \frac{\beta^2}{N}\left(\langle \mathcal{H}^2 \rangle - \langle \mathcal{H} \rangle^2\right), \tag{4}$$

where $\beta = 1/(k_B T)$ and $N$ is the number of pseudospins contributing to the specific heat. The multi–dimensional integrals in the expression for the average is computed by means of the Monte-Carlo (MC) method using Wolff's cluster algorithm [19]. We carried out of the order of $20,000$ thermalization steps and of the order of $750,000$ measurements. The calculations were performed on a heterogeneous environment of computers including Sun,

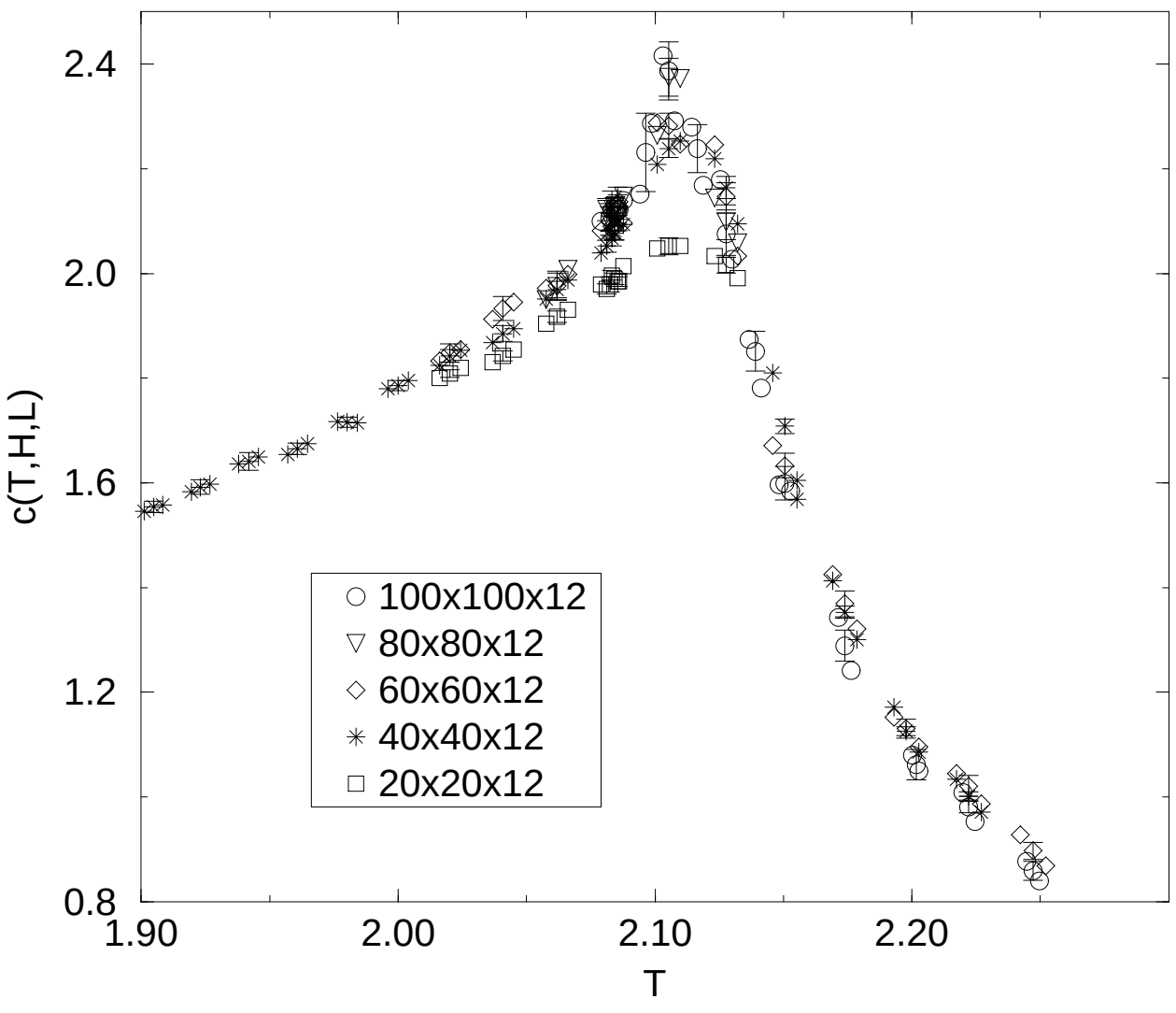

FIG. 1. The specific heat as a function of $T$ for $L^2 \times 12$ lattices. In the 3D $x-y$ model $T_\lambda/J = 2.2017$ [17].

IBM RS/6000 and DEC alpha AXP workstations and a Cray–YMP.

Since we do not possess an easy procedure to extrapolate to the $L \to \infty$ limit for the values of the specific heat computed on finite lattices $L \times L \times H$, we shall use our largest value of $L = 100$ to represent the infinite $L$ limit. This can be justified because the specific heat appears independent of $L$ for $L \geq 60$ (cf. Fig.1). Furthermore, for given thickness $H \leq 24$ used in this work, we do not expect the maximum of the specific heat to grow significantly with increasing values of $L > 100$ because for temperatures in the range $T_c^{2D}(H) \leq T \leq T_\lambda$ (where $T_c^{2D}(H)$ is the Kosterlitz-Thouless critical temperature for films of thickness $H$) the behavior of the specific heat can be described by Kosterlitz–Thouless theory which leads to a finite value of this maximum. In order to illustrate this argument further, we show in Fig.2 the size dependence of the specific heat $c(T, L)$ computed on pure two–dimensional lattices $L \times L$ (up to $L = 400$) with periodic boundary conditions. The $L$–dependence of the specific heat can be neglected for values of $L > 80$.

Now, we would like to check the finite–size scaling hypothesis for the specific heat $c(t, H)$ for large $L$ ($L \gg H$). We consider the finite–size scaling expression for the specific heat $c$ given by [5,6]:

$$c(t, H) = c(t_0, \infty) + H^{\alpha/\nu} f_1(t H^{1/\nu}). \tag{5}$$

The function $f_1(x)$ is universal and $\nu = 0.6705$ as has been extracted from recent accurate experiments [16]. The hyperscaling relation $\alpha = 2 - 3\nu$ yields $\alpha/\nu = -0.0172$. At the reduced temperature $t_0$ the correlation length $\xi(t) = \xi_0^{\pm}|t|^{-\nu}$ becomes equal to the film thickness

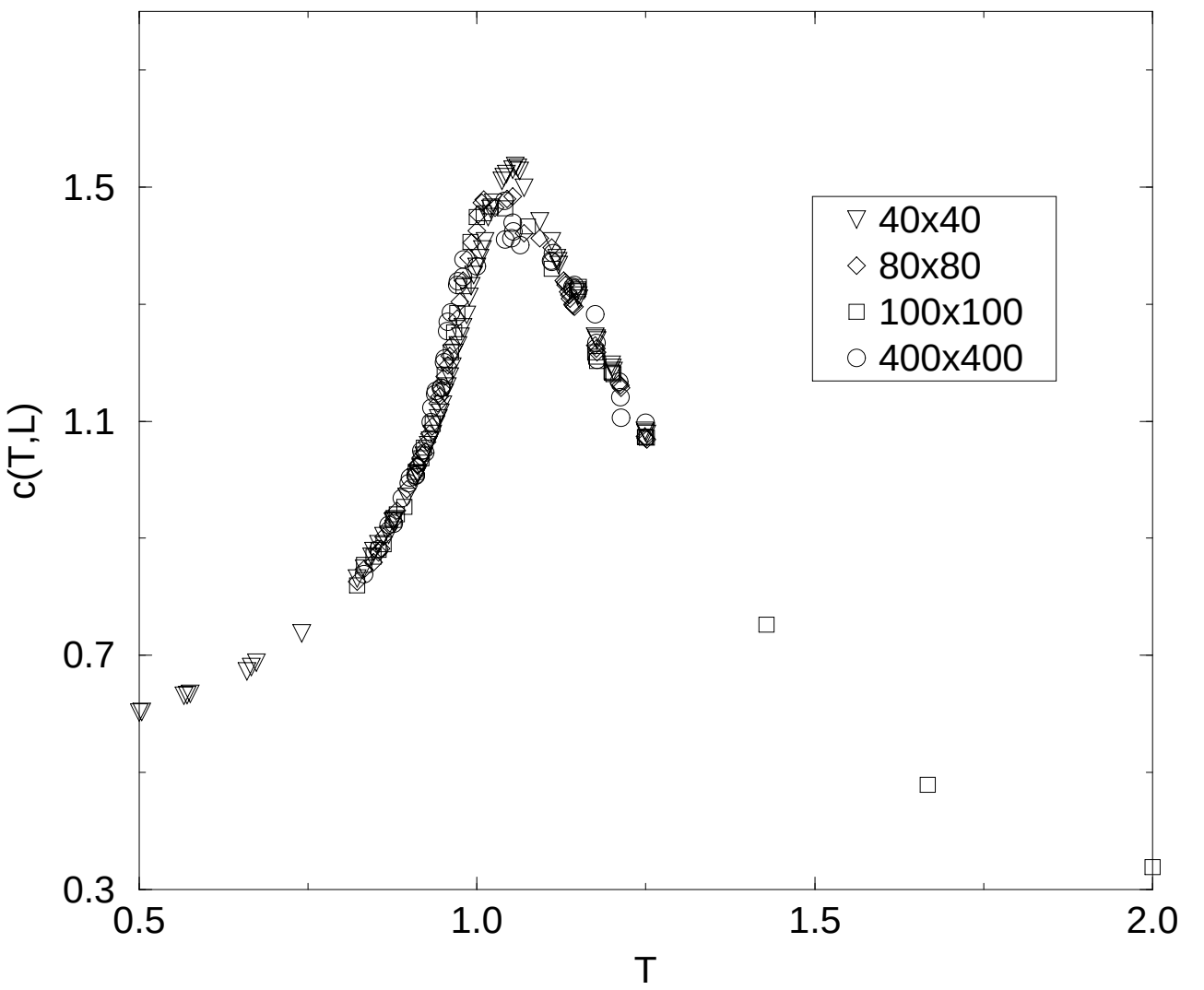

FIG. 2. The specific heat for pure two–dimensional lattices $L \times L$ with periodic boundary conditions.

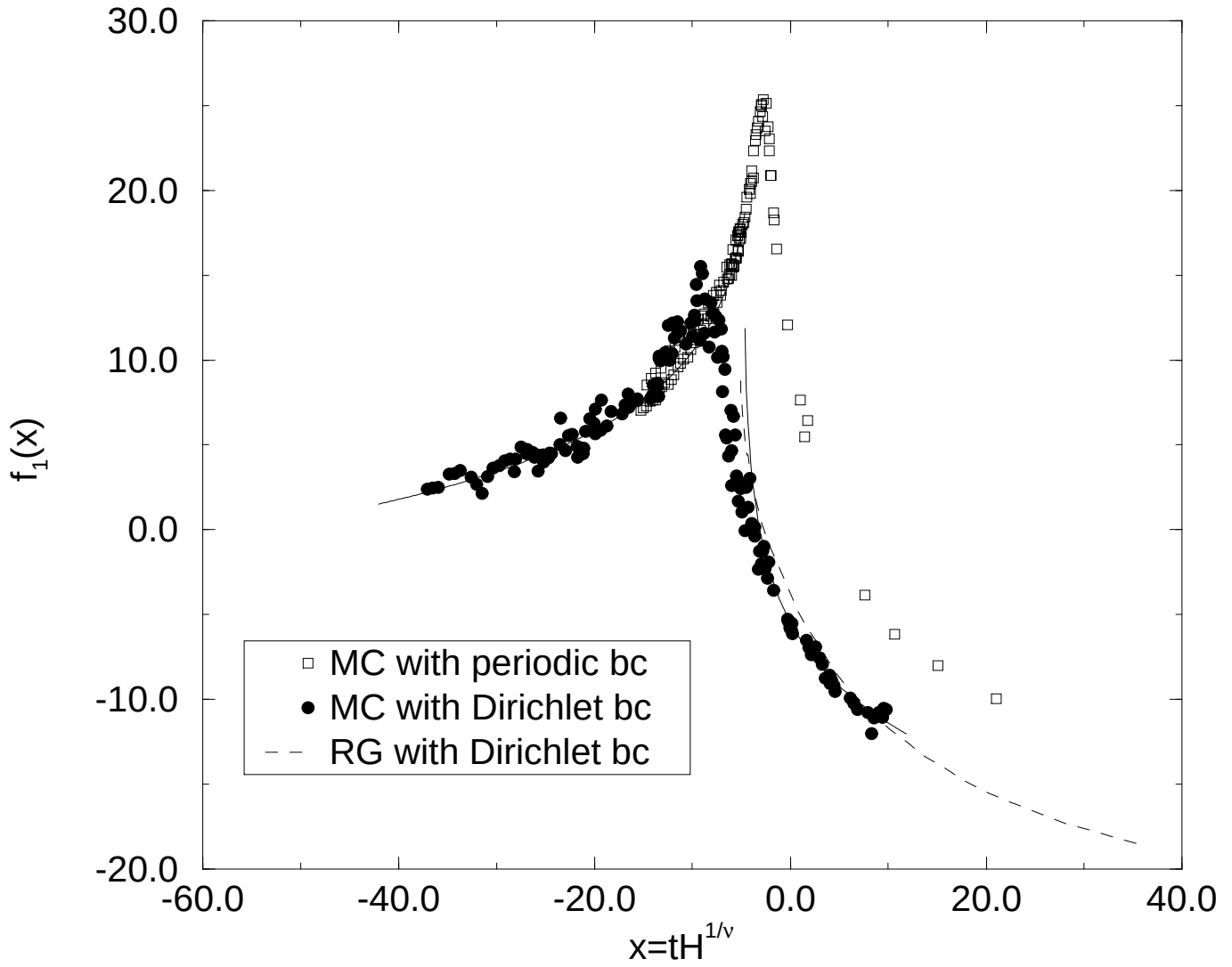

FIG. 3. The universal function $f_1(x)$ for films with Dirichlet boundary conditions is compared to the results of our earlier calculation using periodic boundary conditions [13] (open squares). The solid lines are the results to our fit to the function given by Eq. (10). The dashed line represents the results of the loop expansion calculation [5,6].

$H$, i.e. $t_0 = (\xi_0^+/H)^{1/\nu}$ with $\xi_0^+ = 0.498$ [18]. We have

$$c(t_0, \infty) = c(0, \infty) + \tilde{c}_1^+ t_0^{-\alpha}, \qquad (6)$$

where we use the bulk values $c(0,\infty) = 30$, $\tilde{c}_1^+ = -30$ obtained by studying the finite-size scaling of the specific heat of cubes [13].

Thus, we can make a direct comparison of our universal function $f_1(x)$ to the experimentally determined scaling function $f_1(x)$ given in Refs. [9]. In order to do that we express all lattice units in physical units leading to the conversion formula:

$$f_1(x)|_{phys} = \frac{V_m k_B}{a^3} \left(\frac{a}{\text{Å}}\right)^{-\alpha/\nu} f_1(x)|_{lattice} \qquad (7)$$

where $V_m$ is the molar volume of $^4He$ at saturated vapor pressure at $T_\lambda$. The unit of length $a$ (i.e., the lattice spacing $a$ in the $x-y$ model) can be determined in such a way that the $x-y$ model describes the fluctuations of the order parameter in liquid helium. This means that the calculated bulk helicity modulus $\Upsilon$ from the $x-y$ model (which is related to the superfluid density $\rho_s$ by $\rho_s = (m/\hbar)^2\Upsilon$ where $m$ is the bare $^4He$ atom mass) and the one measured experimentally match near the critical point. Namely, $\beta\Upsilon$ has dimensions of inverse length and near $T_\lambda$ can be expressed in units of $a$ as $\beta\Upsilon(t \to 0) = Y_0 t^\nu / a$ which is matched with the experimental temperature dependence to obtain the value of $a$ in Angstroms. We find $a = 2.95$Å [13] and thus

$$f_1(x)|_{phys} = 15.02\, f_1(x)|_{lattice}\, Joule/(^\circ Kmole). \qquad (8)$$

Therefore, the function $f_1(x)$ in physical units is completely determined.

In Fig.3 we give the results of our present Monte Carlo calculation with Dirichlet boundary conditions along the film thickness direction and we compare them to the results of our earlier Monte Carlo calculation [13] using periodic boundary conditions. Each set of our Monte Carlo data collapse onto a universal curve which defines the function $f_1(x)$ for periodic and a different function $f_1(x)$ for Dirichlet boundary conditions. As a comparison we also plot the result for $f_1(x)$ of the field theoretical calculation for the Landau–Ginzburg functional in a film geometry with Dirichlet boundary conditions [5,6] (dashed line).

The universal function $f_1(x)$ satisfies the following limits

$$f_1(x \to \pm\infty) \to c_\pm |x|^{-\alpha}. \qquad (9)$$

For sufficiently large values of $|x|$ the universal function $f_1(x)$ for the film geometry should agree with the universal function obtained for the cubic geometry (bulk). In our earlier work [13] we found that for cubes $c_+ = -449.6 Joules/(^\circ Kmole)$ and $c_- = -430.8 Joules/(^\circ Kmole)$ for $\alpha = -0.0115$. Using these asymptotic constraints on the form of the function $f_1(x)$ we were able to fit our results for $f_1(x)$ for intermediate and large values of $|x|$ with the formula

$$f_1(x) = c_\pm^0 + c_\pm |x + x_\pm^0|^{-\alpha}. \qquad (10)$$

where the + and − refer to the form which should be used for values of $x$ above and below the peak of the function $f_1(x)$ respectively. By fitting our results to the above function using the above mentioned values of $c_\pm$ we find that $c_+^0 = 452.3 \pm 0.3 Joules/(^\circ K mole)$ and $c_-^0 = 450.8 \pm 0.2 Joules/(^\circ K mole)$ and $x_+^0 = 4.8 \pm 0.4 \mathring{A}$ and $x_-^0 = 3.4 \pm 0.5 \mathring{A}$. Notice that the asymptotic form of Eq. (10) and the limits given by Eq. (9) agree. The asymptotic forms (10) describe our data accurately for $-40 \le x \le -16$ and $-2 \le x \le 10$.

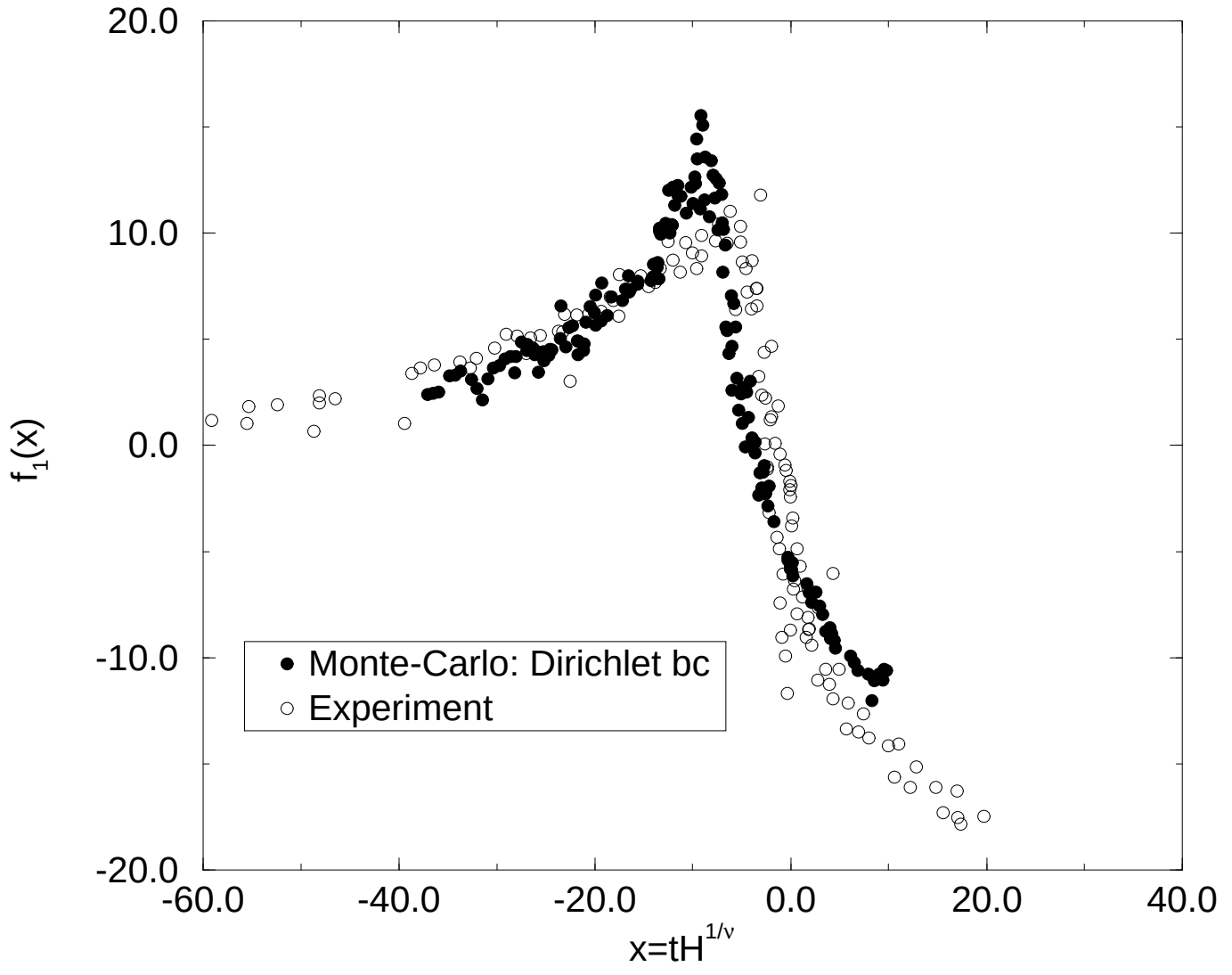

FIG. 4. The universal function $f_1(x)$ for films with Dirichlet boundary conditions is compared to the experimental results [9].

In Fig.4 we show $f_1(x)$ obtained from our Monte–Carlo calculation (solid circles) using Dirichlet boundary conditions and the experimental data points taken from Ref. [9]. The agreement between theory and experiment is quite satisfactory.

The range of the variable $x = tH^{\frac{1}{\nu}}$, in which our calculation of the scaling function $f_1(x)$ is valid, is determined by the requirement $\xi \gg a$ (the lattice spacing) for critical (continuum limit) behavior, which implies that $t \ll ([\xi_0])^{\frac{1}{\nu}}$ ($[\xi_0]$ is the prefactor of the correlation length $\xi(t) = a[\xi_0]t^{-\nu}$ near the lambda point in lattice spacing units) and therefore $|x| \ll ([\xi_0]H)^{\frac{1}{\nu}}$. Taking the numerical values for $[\xi_0] \sim 1.2$ and $H \sim 20$ used in our calculations we find that we can expect scaling for $|x| \ll 100$. Indeed we find that for the values of $H$ used in our calculations we obtain scaling for $|x| < 40$. For $|x| > 40$ we find deviations from scaling. If we wish to obtain the scaling function for even larger values of $|x|$, we need to calculate the specific heat on even larger thickness helium films. However, the calculation of the specific heat for several temperature values on lattices of sizes $100 \times 100 \times H$ with a given value of $H$, e.g. $H = 24$, using cluster Monte Carlo, takes approximately 200 hours of Cray-YMP CPU time; thus, calculations for values of $H$ much larger than the currently studied are beyond realistic computational time scales.

The results of our calculations can be used in a straightforward way to deduce the specific heat values for any size helium films. Such a simple calculation might be useful for the preparation of the ground-based work for the confined helium experiment (CHeX) which is planned to be conducted in 1997 in space under microgravity conditions [11] where the rounding of the transition due to gravity can be avoided.

This work was supported by the National Aeronautics and Space Administration under grant no. NAGW-3326.